\documentclass[sigconf]{acmart}

\renewcommand\footnotetextcopyrightpermission[1]{} 
\usepackage{cleveref}
\usepackage{algorithm}
\usepackage{algorithmic}
\usepackage{subfigure}

\begin{document}

\title{Integrating Active Learning in Causal Inference with Interference: \\
           A Novel Approach in Online Experiments}

\author{Hongtao Zhu}
\affiliation{%
  \institution{National University of Singapore}
  \city{Singapore}
  \state{Singapore}
  \country{Singapore}}
  \email{hongtao.zhu@u.nus.edu}

\author{Sizhe Zhang}
\affiliation{%
  \institution{Tencent}
  \city{Singapore}
  \state{Singapore}
  \country{Singapore}}
  \email{}

\author{Yang Su}
\affiliation{%
  \institution{Tencent}
  \city{Palo Alto}
  \state{California}
  \country{USA}}
  \email{yaangsu@global.tencent.com}

\author{Zhenyu Zhao}
\affiliation{%
  \institution{Tencent}
  \city{Palo Alto}
  \state{California}
  \country{USA}}
  \email{}

\author{Nan Chen}
\affiliation{%
  \institution{National University of Singapore}
  \city{Singapore}
  \state{Singapore}
  \country{Singapore}}
  \email{isecn@nus.edu.sg}

\renewcommand{\shortauthors}{Zhu et al.}

\begin{abstract}
  In the domain of causal inference research, the prevalent potential outcomes framework, notably the Rubin Causal Model (RCM), often overlooks individual interference and assumes independent treatment effects. This assumption, however, is frequently misaligned with the intricate realities of real-world scenarios, where interference is not merely a possibility but a common occurrence. Our research endeavors to address this discrepancy by focusing on the estimation of direct and spillover treatment effects under two assumptions: (1) network-based interference, where treatments on neighbors within connected networks affect one's outcomes, and (2) non-random treatment assignments influenced by confounders. To improve the efficiency of estimating potentially complex effects functions, we introduce an novel active learning approach: Active Learning in Causal Inference with Interference (ACI). This approach uses Gaussian process to flexibly model the direct and spillover treatment effects as a function of a continuous measure of neighbors' treatment assignment. The ACI framework sequentially identifies the experimental settings that demand further data. It further optimizes the treatment assignments under the network interference structure using genetic algorithms to achieve efficient learning outcome.  By applying our method to simulation data and a Tencent game dataset, we demonstrate its feasibility in achieving accurate effects estimations with reduced data requirements. This ACI approach marks a significant advancement in the realm of data efficiency for causal inference, offering a robust and efficient alternative to traditional methodologies, particularly in scenarios characterized by complex interference patterns.
\end{abstract}





\maketitle

\section{Introduction}
\subsection{Motivation}
In the realm of causal inference, observational studies have traditionally relied on regression-based and matching-based approaches to estimate treatment effects. They work with messy data and seek to estimate treatment effects using regression models \cite{huang2022robust,chernozhukov2018double,ma2020robust,li2018balancing} or matching methods \cite{abadie2002simple,abadie2011bias}. Most studies assumed the absence of interference between two individuals, meaning that one individual's information, including treatment, features, and outcomes, does not affect others. In reality, interference is prevalent. For example, one person's increased online gaming time may be influenced by their friends' participation in certain in-game activities. 
Studies in other domains, including but not limited to health \cite{halloran1991study,halloran1995causal}, education \cite{hong2006evaluating,graham2010measuring,vanderweele2013mediation}, social interactions \cite{eckles2016estimating}, and politics \cite{bowers2013reasoning} have shown interference affects treatment effects. Clearly, considering interference in causal inference is a highly practical and relevant topic. 

\subsection{Related work}

Presently, numerous studies within the field of causal inference aim to estimate treatment effects through experimental or non-experimental approaches. The majority of these investigations are grounded in the potential outcomes framework \cite{rubin1974estimating,cox1958planning}, commonly referred to as the Rubin Causal Model (RCM), which is underpinned by three core assumptions: Stable Unit Treatment Value Assumption (SUTVA)  \cite{rubin1980randomization}, Exchangeability Assumption and	Positivity Assumption \cite{rosenbaum1983central,hernan2010causal}. Based on this framework, numerous methodologies for effect estimation in causal inference have been developed \cite{robins1994estimation,li2018balancing,chernozhukov2018double,abadie2002simple,abadie2011bias,yang2023multiply}. Notably, all the methods discussed earlier are based on the assumption of no interference (SUTVA), which posits that one individual's information, including treatment, features, and outcomes, does not affect others. However, interference is a common occurrence in the real world, where one individual's actions or treatment can impact others' outcomes. This real-world interference adds complexity to causal inference and often necessitates the use of advanced statistical and modeling techniques.

Recently, the number of papers addressing causal inference with interference or estimating treatment effects in the presence of interference has grown significantly. There is a growing emphasis on estimating two distinct effects: the direct treatment effect, which refers to the impact of an individual's own treatment on their outcome, and the spillover effect, alternatively termed the indirect treatment effect, which pertains to the influence of treatments assigned to an individual's neighbors on the individual's outcome \cite{hudgens2008toward}.

Current methods for causal inference in the presence of interference can be broadly categorized into two main groups: randomized experiments and observational studies. Randomized experiments methods employ specific randomization techniques, such as cluster randomization or network randomization \cite{hudgens2008toward,baird2014designing,aronow2017estimating,rosenbaum2007interference,eckles2016estimating,bowers2013reasoning,sobel2006randomized}. Early methods were developed to estimate the overall treatment effects in the presence of interference, which is the sum of both the direct treatment effects and the spillover effects \cite{rosenbaum2007interference}. A two-stage randomization procedure has been proposed to facilitate the estimation of direct treatment effects and spillover effects separately within clusters \cite{hudgens2008toward}. There have also been refinements in controlling the variance of treatment effects to obtain better Horvitz–Thompson type estimators \cite{aronow2017estimating}. Additionally, complex scenarios have been introduced, for example, \cite{eckles2016estimating} uses Encouragement Designs to explore the endogenous behavior of interest. However, these methods cannot be applied in observational studies or in specific experiments where proper randomization is not feasible.

In terms of observational studies, some research focuses on the intricacies of causal and network structures, providing direction and theoretical foundation for the further estimation of treatment effects. Notably, studies utilizing Directed Acyclic Graphs (DAGs) have been instrumental in elucidating the causal structure in the presence of interference. For example, works by Ogburn and Sherman present a potential outcomes framework under interference, revealing the intricate structure of interference and guiding further statistical analysis \cite{ogburn2014causal,sherman2018identification}. The following Figure~\ref{fig_c}, as illustrated in \cite{ogburn2014causal}, displays the causal structure in the scenario where interference is present. With a clear definition of causal structure and effects, several methodologies \cite{liu2016inverse,tchetgen2012causal,van2014causal, ogburn2022causal,forastiere2021identification, forastiere2022estimating,leung2022unconfoundedness} have been developed to effectively estimate potential outcomes or effects under various treatment levels.

\begin{figure}[ht]
\begin{center}
\centerline{\includegraphics[width=80pt]{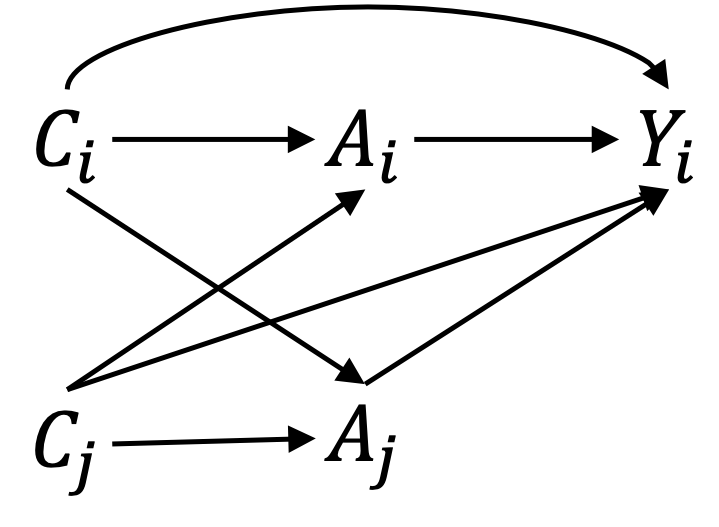}}
\caption{Illustrative causal structure of individual \( i \) interfering with individual \( j \).}
\label{fig_c}
\end{center}
\vskip -0.2in
\end{figure}

The use of Inverse Probability Weighting (IPW) methods has also been extended to address network inference \cite{liu2016inverse,tchetgen2012causal}. \cite{liu2016inverse} build upon the work of \cite{tchetgen2012causal} to introduce a robust IPW estimator to estimate effects based on clusters. However, similar to traditional IPW estimation, this method heavily relies on correct specification of propensity score functions and assumes treatment assignments through Bernoulli allocation strategies. It could lead to significant biases in treatment effect estimation if the model structure of the propensity score functions is misspecified.

\begin{figure}[ht]
\begin{center}
\centerline{\includegraphics[width=200pt]{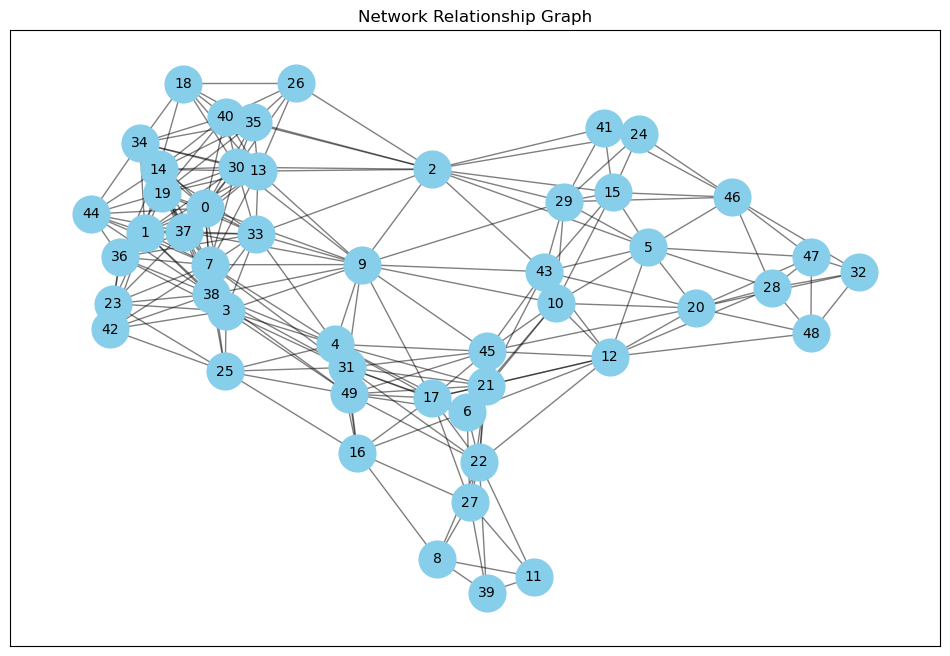}}
\caption{The social network of 50 individuals.}
\label{network}
\end{center}
\vskip -0.2in
\end{figure}

Another branch of methodologies has been developed focusing on network-based approaches (a network example is shown in Figure ~\ref{network}). These include the Targeted Maximum Likelihood Estimator (TMLE) \cite{van2014causal, ogburn2022causal}, methods within the Bayesian framework \cite{forastiere2021identification, forastiere2022estimating}, and applications of Graph Neural Networks (GNNs) \cite{leung2022unconfoundedness}.

The Targeted Maximum Likelihood Estimator (TMLE), as proposed in \cite{van2014causal}, has been utilized in observational studies within networks to estimate average potential outcomes in the presence of interference \cite{ogburn2022causal}. In their research, \cite{ogburn2022causal} employed an approach that integrates the covariates and treatments of neighbors. However, TMLE remains a parametric method, relying on correct model specification. Its primary objective is to estimate the average potential outcome under a determined treatment assignment strategy, which differs from the focus of our research.

The Bayesian framework has been applied to address interference, allowing for more flexible modeling assumptions and complex model structures \cite{forastiere2021identification,forastiere2022estimating}. This approach is also network-based and involves integrating information from neighbors. The estimand in their work aligns with ours, focusing on average effect functions with the integrated treatment level as input. However, a significant challenge arises due to the "Curse of Dimensionality" in the context of high-dimensional covariates. Sparse data in such scenarios tend to skew results towards the prior, impacting the robustness and reliability of the outcomes.

Machine learning techniques, such as deep learning, are also employed to estimate interference effects. Graph Neural Networks (GNNs) have been proposed as a potential solution to the causal inference with interference problem \cite{leung2022unconfoundedness}. A key advantage of GNNs is their ability to circumvent the loss of information often encountered when integrating neighbors' data. However, it is worth noting that GNNs underperform when dealing with overly complex social networks.

In summary, current observational studies in causal inference with interference fall short in effectively addressing the challenges in estimating treatment effects. The most common issue is they rely on correct parametric model assumptions, making it difficult to adapt to the complex real-world problems. 


\subsection{Contributions}

In this paper, we consider the estimation of direct treatment effects and spillover effects when network-based interference is present in online experiments. Instead of relying on parametric model specification, we propose a nonparametric method based on Gaussian process (GP) to quantify the effects at different degree of the interference.  

GP is a flexible regression model that can approximate a wide range of functional relationships. In addition, GP provides uncertainty quantification and points out the areas that require more samples to improve the approximation accuracy. This feature makes it particularly useful in online experiments, where sequential experimental designs are relatively easier to conduct. Hinging on this feature, we propose an active learning strategy for designing and analyzing online experiments. Unlike experiments without interference, where the treatment assignment can be precisely conducted, the degree of interference can hardly be controlled precisely in networks. Therefore, even with active learning in choosing the experimental design, we still need to integrate methods from observational studies to utilize all samples best.  

\begin{figure}[ht]
\begin{center}
\centerline{\includegraphics[width=180pt]{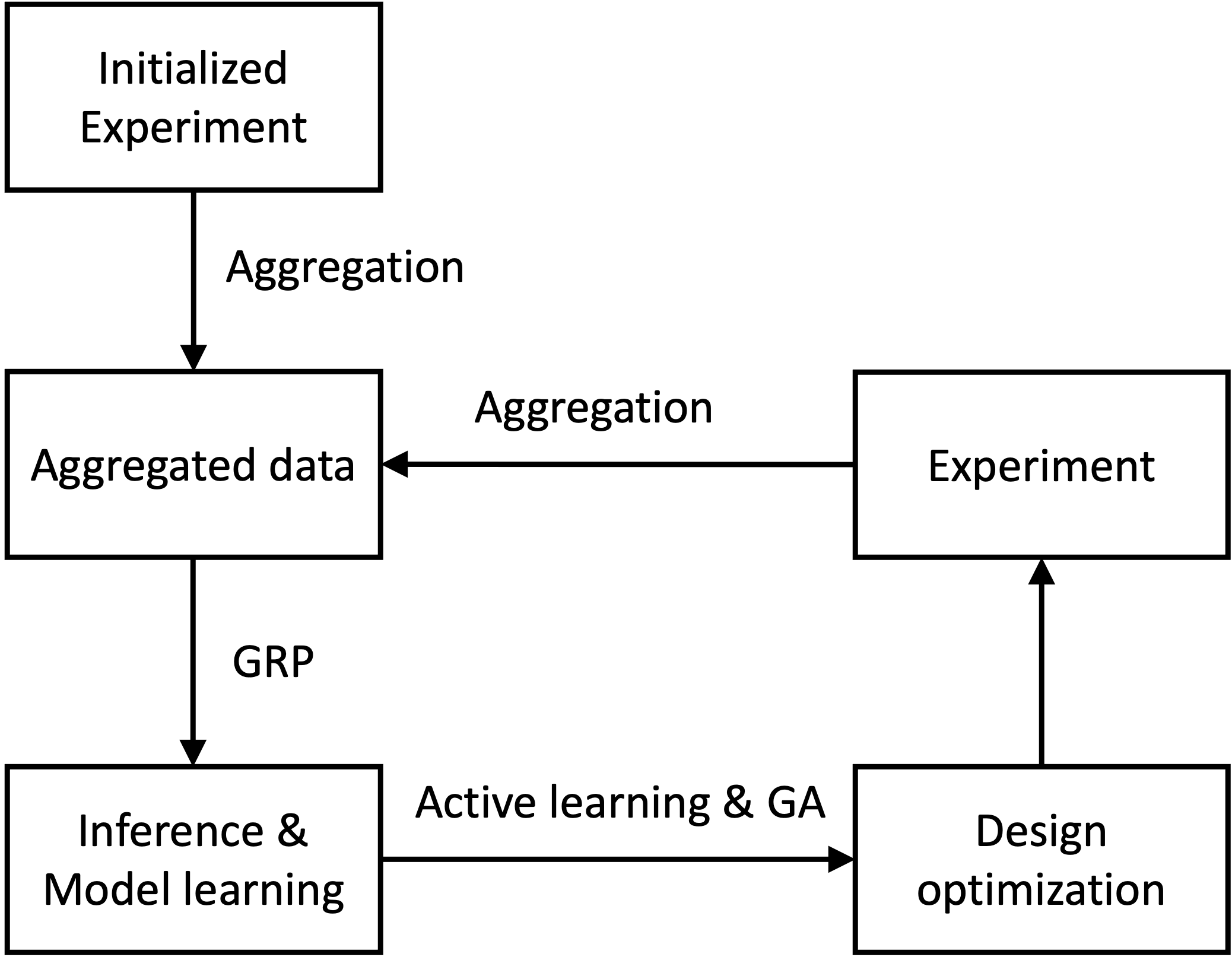}}
\caption{The structure of our active learning framework designed for causal inference with interference (GA: Genetic algorithm, GRP: Gaussian process regression). A detailed exposition of this framework is provided in Section~\ref{sec:proposed_method}.}
\label{fig1}
\end{center}
\vskip -0.2in
\end{figure}

In our proposed framework, we use GP to model the potential outcomes of each individual, given the treatment and covariates of the individual and its  neighbors. Because of the varying network structure, the number of neighbors or the degree of interference can be different. We use active learning to determine the design of new experiments so that we can have data at desired interference levels. This iterative process continues until the budget runs out or the effects have been estimated with sufficient accuracy. The diagram of the iterative process is illustrated in Figure~\ref{fig1}.

In summary, our work's contributions include: (1) We introduce active learning to causal inference with interference, by which we dynamically select the most informative data points to enhance efficiency and accuracy in effect estimation. (2) We employ a non-parametric method for effect estimation, which is resistant to model misspecification and adept at capturing non-linear relationships. (3) We proposed a genetic algorithm to optimize the treatment assignments to have the desired level of network interference. 

The remainder of this article is organized as follows. Section \ref{sec:Preliminaries and Problem Formulation} presents the model formulation and introduces the potential outcomes framework that takes into account network-based interference. Section \ref{sec:proposed_method} discuss in details our methodology, including GP modeling, active learning strategy, and treatment assignment optimization. Section \ref{sec:Experimental Results} demonstrates the performance of the method through simulations and an empirical study using a real dataset.  Section \ref{sec:conclusion} concludes the paper with discussions. 


\section{Preliminaries and Problem Formulation}
\label{sec:Preliminaries and Problem Formulation}

In this section, we outline the key assumptions relevant to our study and provide a comprehensive methodology for integrating covariates and treatments.

\subsection{Network assumption}

\begin{figure}[ht]
\begin{center}
\centerline{\includegraphics[width=0.8\columnwidth]{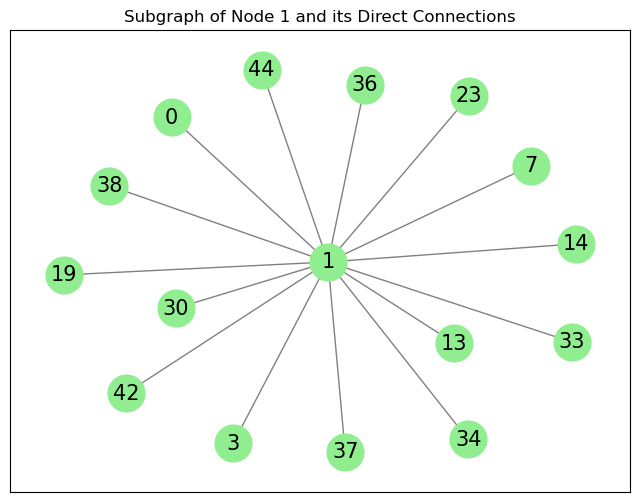}} 
\caption{Illustrative Sub-Network Centered Around Individual \( 1 \), where connections indicate relationships between individuals, e.g., \( w_{1,0} = 1 \).}
\label{fig2}
\end{center}
\end{figure}

To represent networks for all \( n \) individuals, we utilize a relationship matrix denoted as \( W \in \mathbb{R}^{n \times n} \). In this matrix, vector \(W_i\) represents the relationships of individual \(i\) with all its neighbors, and element \( w_{ij} \) indicates the strength of the relationship between two individuals, \( i \) and \( j \). When \( i \) and \( j \) are not neighbors---implying no connection such as in-game friends---the value \( w_{ij} \) is set to 0. In contrast, if \( i \) and \( j \) are neighbors, \( w_{ij} \) is assigned a positive real number reflecting the strength of their relationship. As a special case, \( W \in \mathbb{R}^{n \times n} \) takes binary values, with \( w_{ij} = 1 \) denoting the existence of the relationship between \( i \) and \( j \), and 0 otherwise.  Figure~\ref{fig2} displays a sub-network of the network presented in Figure~\ref{network} for individual $1$. This sub-network includes individual $1$ and all its neighbors, with individual $1$ defined as the central node of this specific sub-network.

\subsection{Potential outcomes framework with interference}

Building upon \cite{rubin1974estimating} and \cite{ogburn2014causal}, we define our potential outcomes framework with interference as follows. In our notation, we employ \( Y \) to denote the outcome, \( A \) to represent a binary treatment, i.e., \( A = 0 \) or \( 1 \), and \( C \) to represent confounders with dimension \( p \). An individual \( i \) is completely characterized by \( \{Y_i, A_i, C_i\} \).

We denote the number of neighbors of individual \( i \) as \( m_i \), and neighbors of individual \( i \) belongs to the collection \( N_i^f = \{ j \mid w_{ij} > 0 \} \). For notation simplicity, we group the data for all \( m_i \) neighbors as \( Y_i^N = [Y_j \mid j \in N_i^f ] \in \mathbb{R}^{m_i} \), \( A_i^N = [A_j \mid j \in N_i^f ] \in \mathbb{R}^{m_i} \), \( C_i^N = [C_j \mid j \in N_i^f ] \in \mathbb{R}^{m_i \times p} \). Throughout this article, we use uppercase letters to indicate random variables, and lowercase letters to indicate a realization, unless otherwise specified.

Each central individual often possesses its unique neighbor structures. For instance, the number of neighbors for individual \( i \) might differ from that of individual \( j \). Consequently, the treatments \( (a_i, a_i^N) \) and \( (a_j, a_j^N) \) may have distinct dimensions and levels. To tackle this issue, a transformation function \( g_i\) is introduced:
\begin{equation}
    G_i = g_i (A_i^N, W_i) 
\end{equation}
Here, \( G_i \) can represent the proportion of neighbors who have been subject to the treatment, or equivalently, the proportion of \( 1 \) within \( A_i^N \). This approximation is similar to the one used by \cite{forastiere2021identification,forastiere2022estimating}.

It is essential to acknowledge that varying network structures can lead to differences in the sizes of \( C_i^N \) and \( C_j^N \). Furthermore, the size of \( C_i^N \) can become considerably large, given the substantial number of neighbors. Considering that individuals receiving different treatments in observational studies are likely to have different distributions in covariates, we aim to control the transformed neighbor covariates within the same and reasonable size while minimizing information loss during the approximation process. To clarify further, our objective is to approximate neighbor covariates in a manner that retains, to a certain extent, their distribution information within specific \( A_i^N \) conditions. To achieve this, we introduce a transformation function \( h_i \):
\begin{equation}
    \tilde{C}_i^N = h_i (C_i^N, A_i^N, W_i) \
\end{equation}
For example, \( \tilde{C}_i^N \) could represent the average of \( C_i^N \), i.e., 
\[
\tilde{C}_i^N = \frac{\sum_{j=1}^{m_i} C_{i,j}^N}{m_i},
\]
where \( m_i \) can be any natural number (\( m_i \in \mathbb{N} \)), accounting for the variable number of neighbors among central individual \( i \). Of course, other reasonable choices of $g_i, h_i$ functions can be used. 

Considering the neighbors' information, an individual \( i \) can be characterized by \( \{Y_i, (A_i, G_i), X_i\} \)
where \( X_i = (C_i, \tilde{C}_i^N) \). The three causal assumptions from section 2.2 can be expressed as follows:
\begin{itemize}
    \item Assumption of consistency under interference:
    \begin{equation}
        Y_i(a, g) = Y_i \quad \text{when} \quad (A_i, G_i) = (a, g) 
    \end{equation}
    \item Exchangeability assumption:
    \begin{equation}
        Y_i(a, g) \perp\!\!\!\perp (A_i, G_i) \mid X_i 
    \end{equation}
    \item Positivity assumption:
    \begin{equation}
        P((A_i, G_i) = (a, g) \mid X_i) > 0 
    \end{equation}
\end{itemize}

\subsection{Causal estimands}

When considering the treatment \( (a, g) \) and the controlled level \( (0, 0) \), we define an individual's overall treatment effect (represented as \( \text{IOTE}_i \)), direct treatment effect (represented as \( \text{IDTE}_i \)) and spillover effect (depicted as \( \text{ISE}_i \)) as follows:
\[
    \text{IOTE}_i(g) = \mathbb{E}[Y(1, g) | X_i] - \mathbb{E}[Y(0, 0) | X_i]
\]
\[
    \text{IDTE}_i(g) = \mathbb{E}[Y(1, g) | X_i] - \mathbb{E}[Y(0, g) | X_i]
\]
\[
    \text{ISE}_i(g) = \mathbb{E}[Y(0, g) | X_i] - \mathbb{E}[Y(0, 0) | X_i]
\]

Similarly, we provide average level effects as follows, with average overall treatment effect \(\tau_1\), average direct treatment effect \(\tau_{1,0}\) and average spillover effect \(\tau_0\):
\[
    \tau_1(g) = \mathbb{E}[\mathbb{E}[Y(1,g) | X] - \mathbb{E}[Y(0,0) | X]]
\]
\[
    \tau_{1,0}(g) = \mathbb{E}[\mathbb{E}[Y(1,g) | X] - \mathbb{E}[Y(0,g) | X]]
\]
\[
    \tau_0(g) = \mathbb{E}[\mathbb{E}[Y(0,g) | X] - \mathbb{E}[Y(0,0) | X]]
\]

\section{Proposed Method}
\label{sec:proposed_method}

We divide our proposed method into two parts. First, we use a method based on Gaussian process regression to estimate potential outcomes, thereby estimating individual treatment effects and average treatment effects. Second, we use active learning to identify how to assign new samples so that the treatment effects can be better estimated. 

\subsection{Gaussian Process Regression for Treatment Effects}

 Our proposed approach employs Gaussian Process Regression (GPR) to assess the treatment effects on the entire population at a specific treatment level \( A = a \) and \( G = g \). Specifically, we model the potential outcome as at treatment level $a$ as $f_a(g,X)$. This methodology leverages all available observational data, especially outcomes from experimental settings, when the condition \( A = a \) is met. We aim to forecast the potential outcomes for the whole population under the treatment level \( A = a \) and \( G = g \). 
 
To implement Gaussian Process regression, we define our kernel function as \( k((\boldsymbol{x}, g), (\boldsymbol{x'}, g')) \). For more detailed examples of this kernel function, refer to the Appendix~\ref{subsec:Combined-kernel-function}. Consider a training dataset for a specific treatment \( A = a \), with size \( n_t \). This dataset comprises integrated covariates \( \boldsymbol{X_t} = (X_{t,1}, \ldots, X_{t,n_t})^T \), integrated neighbors' treatments \( \boldsymbol{g_t} = (g_{t,1}, \ldots, g_{t,n_t})^T \), and their actual outcomes \( \boldsymbol{y_{t}} = (y_{t,1}, \ldots, y_{t,n_t})^T \). By following the maximum likelihood framework, we can estimate the hyper-parameters to maximize the marginal (log) likelihood of the training data, as shown in Appendix~\ref{subsec:Optimizing-hyper-parameters}. Our goal is to predict the potential outcomes for the entire population, which includes a total of \( n \) individuals with integrated covariates \( \boldsymbol{X} = (X_1, \ldots, X_n)^T \), under the treatment level \( A = a, G = g \). The predicted outcome function values are denoted as \( \boldsymbol{f_a}(g, \boldsymbol{X}) = [f_a(g, X_1), \ldots, f_a(g, X_n)]^T \).

To estimate the average treatment effects, we should determine the average potential outcome function \( m_a(g) \) for the entire population under the treatment level \( (A = a, G = g) \). This function is expressed as follows:
\begin{equation}
\begin{aligned}
m_a(g) = \mathbb{E}_X[f_a(g, X)] = \int f_a(g, X) \, d\mathbb{P}(X),
\end{aligned}
\end{equation}
which can be estimated by:
\begin{equation}
\begin{aligned}
\hat{m}_a(g) = \frac{1}{n} \sum_{i=1}^{n} f_a(g, X_i) = \frac{1}{n} \boldsymbol{e}^T \cdot \boldsymbol{f_a}(g, \boldsymbol{X}),
\end{aligned}
\end{equation}
where \( \boldsymbol{e} = (1,...,1)^T \) with length \( n \).

We represent the kernel matrix used in our model as follows: \(K(\boldsymbol{X_t},\boldsymbol{g_t}),(\boldsymbol{X_t},\boldsymbol{g_t}))=\boldsymbol{K_t}\), \(K((\boldsymbol{X},\boldsymbol{g}),(\boldsymbol{X},\boldsymbol{g}))=\boldsymbol{K_*}\) and \(K((\boldsymbol{X_t},\boldsymbol{g_t}),\\(\boldsymbol{X},\boldsymbol{g}))=\boldsymbol{K_{t,*}}\), where \( \boldsymbol{g} = (g,...,g)^T \) is a vector with length \( n \). 
It then follows that these potential outcomes adhere to a multivariate normal distribution, characterized by the mean:
\begin{equation}
    \mathbb{E}[\boldsymbol{f_a}(g, \boldsymbol{X})] = \boldsymbol{K_{t,*}^T} \cdot [\boldsymbol{K_t} + \sigma_t^2\boldsymbol{I}]^{-1} \cdot \boldsymbol{y_{t}}
\end{equation}
And covariance matrix:
\begin{equation}
\begin{aligned}
    \mathrm{Cov}[\boldsymbol{f_a}(g, \boldsymbol{X})] &= \boldsymbol{K_*} - \boldsymbol{K_{t,*}^T} \cdot [\boldsymbol{K_t} + \sigma_t^2\boldsymbol{I}]^{-1} \cdot \boldsymbol{K_{t,*}}
\end{aligned}
\end{equation}

Following the definitions of overall treatment effects function and spillover effects function, their estimators can be shown as follows,
\begin{equation}
\begin{aligned}
\hat{\tau}_1(g) = \hat{m}_1(g) - \mathbb{E}[\hat{m}_0(0)]
\end{aligned}
\end{equation}
\begin{equation}
\begin{aligned}
\hat{\tau}_0(g) = \hat{m}_0(g) - \mathbb{E}[\hat{m}_0(0)]
\end{aligned}
\end{equation}

The linear combination of multivariate normal distributions still follows a normal distribution. Therefore, we have,
\begin{equation}
\begin{aligned}
&\hat{\tau}_1(g) \sim N(\mathbb{E}[\hat{m}_1(g)]-\mathbb{E}[\hat{m}_0(g)], Var[\hat{m}_1(g)]) \\
&\sim N\left( \frac{1}{n} (\mathbb{E}[\boldsymbol{f_1}(g, \boldsymbol{X})] - \mathbb{E}[\boldsymbol{f_0}(0, \boldsymbol{X})]), \frac{1}{n^2} (\boldsymbol{e}^T \Sigma[\boldsymbol{f_1}(g, \boldsymbol{X})] \boldsymbol{e}) \right)
\end{aligned}
\end{equation}
\begin{equation}
\begin{aligned}
&\hat{\tau}_0(g) \sim N(\mathbb{E}[\hat{m}_0(g)]-\mathbb{E}[\hat{m}_0(g)], Var[\hat{m}_0(g)]) \\
&\sim N\left( \frac{1}{n} (\mathbb{E}[\boldsymbol{f_0}(g, \boldsymbol{X})] - \mathbb{E}[\boldsymbol{f_0}(0, \boldsymbol{X})]), \frac{1}{n^2} (\boldsymbol{e}^T \Sigma[\boldsymbol{f_0}(g, \boldsymbol{X})] \boldsymbol{e}) \right)
\end{aligned}
\end{equation}

\subsection{Active Learning with Optimal Assignments}
To improve the estimation accuracies, we can iteratively update the experimental design. Our iteration is consist of two steps.  The first step identifies the the experimental condition \((a^*, g^*)\) based on existing estimation. It is expected that additional samples with this setting will improve the overall treatment effects.  In the second step, given the target \((a^*, g^*)\), we choose appropriate networks and optimize the treatment assignments for subsequent  experiment. 
\subsubsection{The selection of the target}

In the initial experiments,  conditions \(A=1, G=1\) and \(A=0, G=0\) should have been included. For network data, these two scenarios are easily achievable. This can be done  by simply setting the treatment for all individuals in the network to either 1 or 0. Consequently, all the data in the current network will satisfy the aforementioned conditions.

The treatment level selection is determined at the location where the variance of the functions $\hat\tau_a(g)$ is maximum. Because of the unique feature of Gaussian process, this variance function is analytically available. This can be mathematically defined as:
\[
a^*, g^* = \underset{a, g}{\mathrm{argmax}}\, \text{Var}[\hat{\tau}_a(g)].
\]
We want to highlight that in active learning literature, other type of criteria can be used as well. For example, choosing the $(a^*, g^*)$ such that the integrated mean square error of the entire function can be reduced most.

\subsubsection{Mapping the pair \((a^*, g^*)\) onto the vector \(\boldsymbol{\mathcal{A}^*}\)}

Upon the selection of the target condition \((a^*, g^*)\), it becomes imperative to determine the assignments for subsequent experiments, thereby facilitating the active learning process. This necessitates the allocation of a vector-based treatment \(\boldsymbol{\mathcal{A}^*}\) within the network. To enhance the efficacy of the matching method in estimating effects, the mapping from \((a^*, g^*)\) to \(\boldsymbol{\mathcal{A}^*}\) aims to achieve two primary objectives: first, it maximizes the volume of data within a small interval surrounding the target \((a^*, g^*)\), which is \((A=a^*, G \in \mathcal{G}^*)\); second it  maximizes the dispersion of the covariates \( X \) within the range \((A=a^*, G \in \mathcal{G}^*)\). (\( \mathcal{G}^* = [g^* - \alpha/2, g^* + \alpha/2] \))

In the realm of applications, our sample data often encompass multiple networks or can be easily segmented into multiple networks, each with distinct configurations and sizes. Specifically, we consider a scenario with \( Q \) networks. The \( q \)-th network, denoted by \( N = N_q \), comprises \( n_q \) individuals. These heterogeneous networks exhibit distinct structural properties, leading to heterogeneity in the transformation from treatment assignment \( \boldsymbol{\mathcal{A}_q} \) to the network interference measure (transformed neighbor treatment) \( G \), as well as in the transformation of covariates within neighboring units. For this network, after being assigned with vector-based treatment \( \boldsymbol{\mathcal{A}_q} \), we have some individuals under treatment with \((A=a^*, G \in \mathcal{G}^*)\). This subset of individuals owns a covariates set \( \boldsymbol{X_{q}^r}(\mathcal{A}_q) = (X_{q,1}^r(\boldsymbol{\mathcal{A}_q}),...,X_{q,n_{q}^r(\boldsymbol{\mathcal{A}_q})}^r(\boldsymbol{\mathcal{A}_q}))^T \) with length \( n_{q}^r(\boldsymbol{\mathcal{A}_q}) \). In order to obtain the optimal \( \boldsymbol{\mathcal{A}_q^*} \), we maximize the fitness function for different networks (by using the genetic algorithm in \cref{alg:optimization_network}) (\cite{kolacz2016measures}),
\vspace{-2.5mm}
\begin{equation}
\begin{aligned}
\max_{\boldsymbol{\mathcal{A}_q} \in \mathbb{R}^{n_q}} \sum_{i=1}^{n_{q}^r(\boldsymbol{\mathcal{A}_q})} \sum_{j=1}^{n_{q}^r(\boldsymbol{\mathcal{A}_q})} d^2(X_{q,i}^r(\boldsymbol{\mathcal{A}_q}), X_{q,j}^r(\boldsymbol{\mathcal{A}_q})),
\end{aligned}
\label{eq:fit}
\end{equation}
where \( d(X_{q,i}^r(\boldsymbol{\mathcal{A}_q}), X_{q,j}^r(\boldsymbol{\mathcal{A}_q})) \) denotes the distance between these two covariates, which could be the Euclidean distance, the Manhattan distance, or other similar distance measures, to describe dispersion while also relating to the number of samples.

Each network (for example, the \( q \)-th network) will obtain an optimal solution \( \boldsymbol{\mathcal{A}_q^*} \), along with the corresponding optimal value \( F_q^* \). The amount of data in the vicinity of the targeted interval around \((a^*, g^*)\) is denoted as \(n_{q}^r(\boldsymbol{\mathcal{A}_q^*})\).
Upon selecting a target \((a^*, g^*)\), to effectively utilize the matching method, we define the lower bound of data volume for \((A = a^*, G \in \mathcal{G^*})\) as \(M\).

In summary, \cref{alg:my_algorithm} shows the Active Learning Algorithm with initialization \(N = \{N_1, \ldots, N_Q\}\), and the dataset \(D = \{X, A, G, Y\}\), which initially starts as a set containing only the data of covariates, with other elements being empty.
Note, in \cref{alg:my_algorithm}, \(D\mid_{(A = a^*, G \in \mathcal{G}^*)}\) denotes the subset of data entries in \(D\) that have been experimented with and stored, satisfying the conditions \((A = a^*, G \in \mathcal{G}^*)\). The notation \(\left| D\mid_{(A = a^*, G \in \mathcal{G}^*)}\right|\) denotes the number of such entries.

\begin{algorithm}
\caption{Active learning Algorithm Description}
\label{alg:my_algorithm}
\begin{algorithmic}
\STATE Input networks \( N \) and dataset \( D \), lower bound \( M \), empty points set \( E \), using treatment levels number upper limit \( T \)
\STATE Get initial targeted treatment level \( (a^*, g^*) \)
\FOR{\( i \) in \( T \)}
    \FOR{\( N_q \) in \( N \)}
        \STATE Get \( \boldsymbol{\mathcal{A}_q^*} \), \( F_q^* \), \( m_q^* \) by fitting the formula \cref{eq:fit} under targeted treatment level \( (a^*, g^*) \)
        \IF{\( n_{q}^r(\boldsymbol{\mathcal{A}_q^*}) \geq M -\left| D\mid _{(A = a^*, G \in \mathcal{G})}\right| \)}
            \STATE \( N = N \setminus N_q \)
            \STATE Do the treatment assignment \( \boldsymbol{\mathcal{A}_q^*} \) on \( N_q \), and add data to \( D \)
        \ENDIF
    \ENDFOR
    \IF{\( n_{q}^r(\boldsymbol{\mathcal{A}_q^*}) \) for all \( N_q \) in \( N \) is smaller than \( M - \left| D\mid _{(A = a^*, G \in \mathcal{G})}\right| \)}
        \STATE Sort \( F^* \) for all networks decreasingly. Get rank \( R \)
        \STATE Get \( k \) with \( \sum_{i=1}^{k-1} F_{R^{-1}(i)}^* < M - \left| D\mid _{(A = a^*, G \in \mathcal{G})}\right| \) and \( \sum_{i=1}^k F_{R^{-1}(i)}^* \geq M - \left| D\mid _{(A = a^*, G \in \mathcal{G})}\right| \)
        \STATE Assign treatment \( \{\boldsymbol{\mathcal{A}_{R^{-1}(1)}^*}, \ldots, \boldsymbol{\mathcal{A}_{R^{-1}(k)}^*}\} \) onto \( \{N_{R^{-1}(1)}, \ldots, N_{R^{-1}(k)}\} \), and add data to \( D \)
        \STATE \( N = N \setminus \{N_{R^{-1}(1)}, \ldots, N_{R^{-1}(k)}\} \)
    \ENDIF
    \STATE Add \( (a^*, g^*) \) to \( E \)
    \STATE Do Gaussian Process Regression to get average overall treatment effect functions and average spillover effect functions
\STATE Select new \( (a^*, g^*) \)
\ENDFOR
\end{algorithmic}
\end{algorithm}

\section{Experimental Results}
\label{sec:Experimental Results}

\subsection{Simulation Data Experiment}

Our simulation experiments utilized 100 networks, each contains 100 individuals with three features. Among these, \(X_1\) is a binary variable taking on values of 0 or 1, while the other two, \(X_2\) and \(X_3\), are continuous variables ranging from 0 to 1. The actual outcome \(Y_i(a, g)\) for each individual under the treatment level \((a, g)\) is given by:
\begin{equation}
\scalebox{0.8}{ 
$\begin{aligned}
&Y_i(a, g) = (\beta_{1,1} \cdot C_{1,i}^2 + \beta_{1,2} \cdot C_{2,i} + \beta_{1,3} \cdot \frac{1}{C_{3,i} + 1}) \cdot a \\
& + (\beta_{0,1} \cdot C_{1,i}^2 + \beta_{0,2} \cdot C_{2,i} + \beta_{0,3} \cdot \frac{1}{C_{3,i} + 1}) \cdot (1 - a) \\
& + \left((\beta_{1,1}^N \cdot (C_{1,i}^N)^2 + \beta_{1,2}^N \cdot \frac{1}{1 + C_{2,i}^N} + \beta_{1,3}^N \cdot C_{3,i}^N) \cdot g \right. \\
& \left. + (\beta_{0,1}^N \cdot (C_{1,i}^N)^2 + \beta_{0,2}^N \cdot \frac{1}{1 + C_{2,i}^N} + \beta_{0,3}^N \cdot C_{3,i}^N) \cdot (1 - g)\right) \\
& \cdot (g - 0.5) + \varepsilon_i,
\end{aligned}$
}
\end{equation}
where the neighborhood averages \(X_{k,i}^N\) for \(k = 1, 2, 3\) are defined as:
\begin{equation}
C_{k,i}^N = \frac{\sum_{j=1}^n w_{ij} \cdot C_{k,j}}{\sum_{j=1}^n w_{ij}}, \quad k = 1, 2, 3.
\end{equation}

\begin{figure}[ht]
    \centering
    \subfigure[Experimented with 4 networks]{%
        \label{fig:first}
        \includegraphics[width=0.45\linewidth]{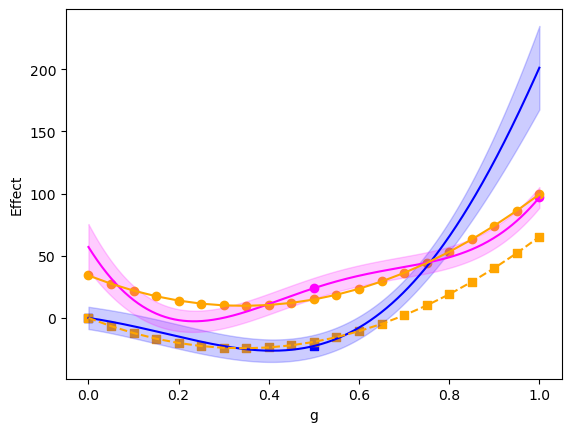}%
    }
    \subfigure[Experimented with 6 networks]{%
        \label{fig:second}
        \includegraphics[width=0.45\linewidth]{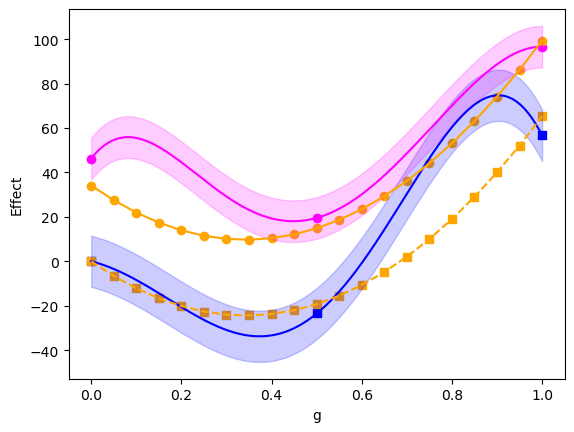}%
    }
    \subfigure[Experimented with 7 networks]{%
        \label{fig:third}
        \includegraphics[width=0.45\linewidth]{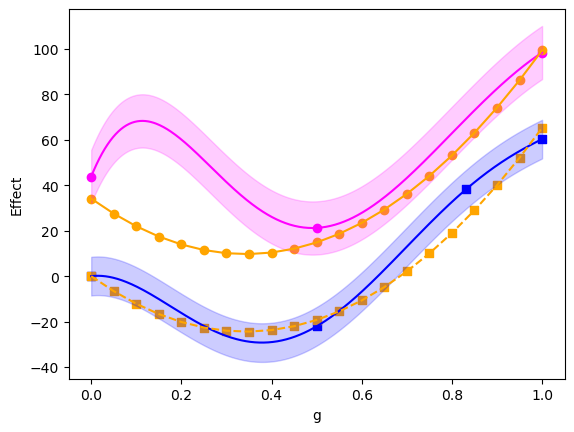}%
    }
    \subfigure[Experimented with 9 networks]{%
        \label{fig:fourth}
        \includegraphics[width=0.45\linewidth]{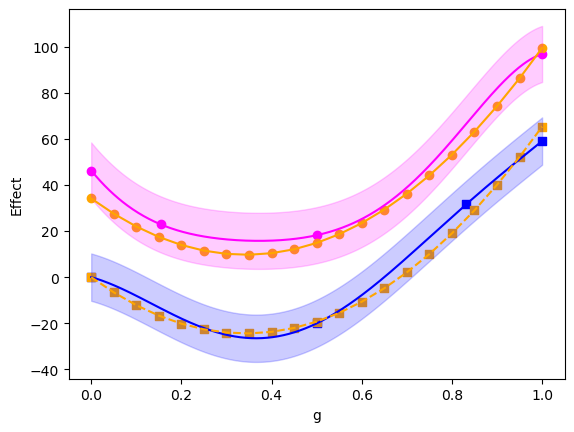}%
    }
    \caption{Active Learning in Causal Inference with Interference (ACI) method: The light pink and dark blue curves represent the estimated average overall and spillover effects, respectively. Points along these curves represent sequentially selected treatment levels. It should be noted that there may be instances where two updates occur simultaneously, yet not all are depicted graphically; for example, transitioning from subfigure (a) to (b) introduces two additional sequential treatment levels. The actual scenarios of these effects are shown by two orange dashed curves. }
    \label{fig:simulation_results}
\end{figure}

Based on our active learning framework applied to this simulation data, we obtained the results shown in Figure~\ref{fig:simulation_results}. Note that after initializing the calculations for the points \((0,0)\) and \((1,1)\), our next choice of points was \((0,0.5)\) and \((1,0.5)\). This intuitive selection aids in our initial Gaussian process regression, after which we adopt a treatment level selection method based on maximum uncertainty.
\begin{figure}
    \centering
    \subfigure[Experimented with 4 networks]{%
        \label{fig:firstc}
        \includegraphics[width=0.45\linewidth]{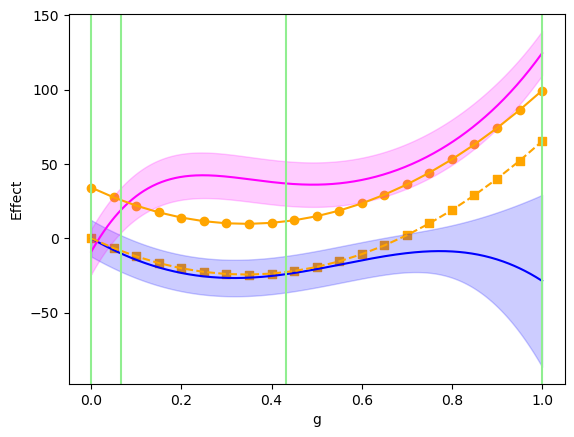}%
    }
    \subfigure[Experimented with 6 networks]{%
        \label{fig:secondc}
        \includegraphics[width=0.45\linewidth]{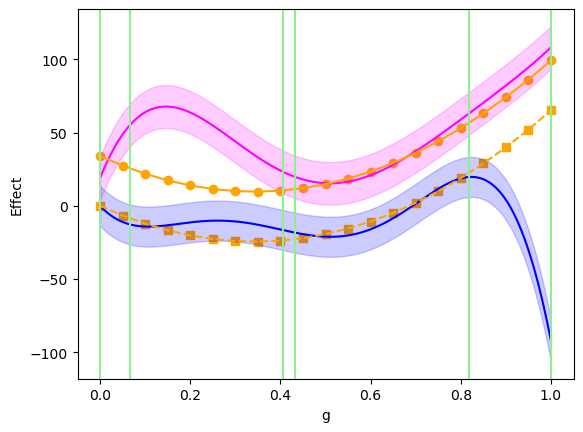}%
    }
    \subfigure[Experimented with 7 networks]{%
        \label{fig:thirdc}
        \includegraphics[width=0.45\linewidth]{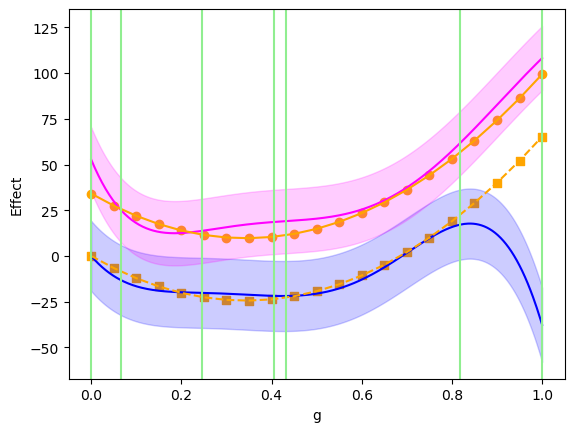}%
    }
    \subfigure[Experimented with 9 networks]{%
        \label{fig:fourthc}
        \includegraphics[width=0.45\linewidth]{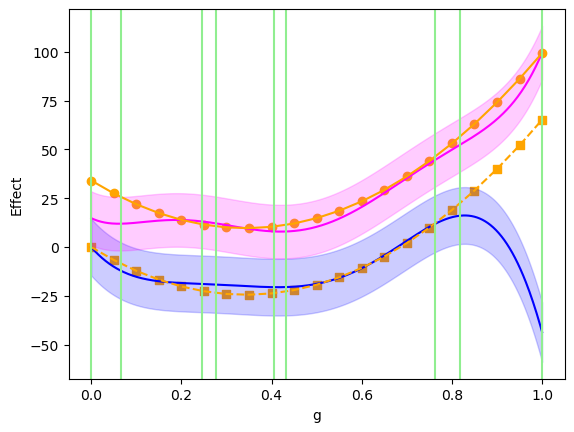}%
    }
    \caption{Comparative random treatment allocation (RTA) method: The light pink and dark blue curves illustrate the regression-derived average overall treatment effect and spillover effect . The light green vertical lines represent the position of the randomly selected \( g^* \).}
    \label{fig:simulation_results_c}
\end{figure}

We observe that in Figure~\ref{fig:fourth}, we have achieved considerably good regression results for both overall treatment effects function and spillover effects function. At this point, we have only used 9 networks, conducting experiments on a total of \(100 \times 9\) individuals. This number is significantly less than the \(100 \times 100 = 10,000\) individuals in the complete dataset. In essence, our method has enabled us to make accurate estimates of the average effects for 10,000 individuals using only a limited number of experiments. 

The simple random treatment allocation (RTA) is applied for comparison. It first  randomly selects the  \( g^* \) for subsequent experiments, and then it also randomly generates the vector-based treatments allocation based on \( g^* \), as detailed in Appendix~\ref{subsec:Random-treatment-allocation}. 
From the simulation comparisons, the proposed ACI method surpass those of the RTA method. Moreover, the ACI method demonstrates a process where a function is sequentially optimized in a data-driven manner. Although the RTA’s predicted functions also show improvement, they reach a certain limit, making it challenging to fit good functions perfectly across the entire range.

Additionally, the Estimated Integral Square Errors (EISE), defined below,  
\[
EISE(\hat{\tau}_1) = \int (\hat{\tau}_1(g) - \tau_1(g))^2 \, d g
\]
\[
EISE(\hat{\tau}_0) = \int (\hat{\tau}_0(g) - \tau_0(g))^2 \, d g
\]
for two different average effects functions at different stages are presented in Table~\ref{tab:EISE_results}.

These superiorities of the ACI method can be attributed to two main reasons: (i) the RTA method lacks a data-driven selection of the optimal treatment level for the next experiment based on existing experimental results; (ii) RTA does not utilize an optimization algorithm to map the target treatment level to an optimal vector-based treatment level, making it challenging to capture the data characteristics at this target treatment level fully. 

\begin{table}[ht]
\centering
\caption{Comparison of EISE for ACI and RTA}
\label{tab:EISE_results}
\begin{tabular}{|c|c|c|c|}
\hline
\textbf{\shortstack{Experimented\\networks'\\ number}} & \textbf{\shortstack{Effect\\ functions}} &\multicolumn{2}{c|}{\textbf{EISE}} \\
\cline{3-4}
&  & \textbf{ACI} & \textbf{RTA} \\ \hline
4 & \(\tau_1(g)\) & 85.39438557 & 447.59047721\\ 
  & \(\tau_0(g)\) & 1814.645103 & 765.83402693\\ \hline 
6 & \(\tau_1(g)\) & 338.79261773 & 565.92932241 \\ 
  & \(\tau_0(g)\) & 395.465169 & 968.10501448 \\ \hline 
7 & \(\tau_1(g)\) & 592.60175094 & 34.5195497\\ 
  & \(\tau_0(g)\) & 51.28109163 & 443.99716071\\ \hline 
9 & \(\tau_1(g)\) & 31.74120205 & 39.42437691\\ 
  & \(\tau_0(g)\) & 14.60991894 & 517.14218562\\ \hline 
\end{tabular}
\end{table}

Moreover, it is important to emphasize that RTA essentially employs the GPR method. Although it cannot guarantee optimal data selection in the active learning phase, as can be observed in Figure~\ref{fig:fourthc}, the predicted functions remain accurate within certain ranges where data is readily available. This further illustrates the advantage of our GPR method. However, to elaborate further, Figure~\ref{fig:simulation_results_c} achieves better results within a certain range largely due to good fortune. Given that our experiments are conducted on a network-level basis, the number of individuals involved in each experiment correlates with the size of the network. This leads to the possibility, under the RTA method, of obtaining a significant amount of data at various treatment levels due to favorable circumstances. Yet, even under these conditions, its results are still inferior to ACI's. Not to mention that, in reality, we might be able to use graph theory knowledge to delineate smaller yet reasonable networks for experimentation.

\subsection{Real Data Experiment}

\begin{figure}[ht]
\begin{center}
\centerline{\includegraphics[width=0.8\columnwidth]{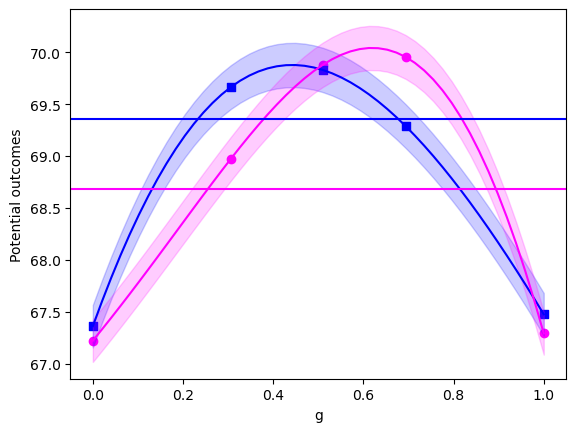}} 
\caption{The case study's implementation of the Active Learning in Causal Inference with Interference (ACI) method: The light pink and dark blue functions represent the estimated average overall and spillover potential outcomes, respectively. Points along these curves represent sequentially selected treatment levels. The horizontal lines in light pink and dark blue, respectively, signify the predicted potential outcomes under treatment and control conditions, excluding the consideration of interference.}
\label{fig:case_study}
\end{center}
\end{figure}

Working in partnership with Tencent, we conducted a practical evaluation of our proposed methodology using data from an online experiment on a mobile game. This gaming experiment aimed to ascertain whether or not a treatment led to increased player engagement compared to the control group. Engagement was measured, for example, by the average online time of the user population. Our analysis investigated the effects of own treatment \(A\) and integrated neighbors' treatment \(G\) on the average online time across the user community. We established relational networks among the participants, considering in-game social connections, and included both the number of friends and the KDA ratio (Kill, Death, Assist ratio, an  common indicator of gaming proficiency) as key covariates. Employing an active learning approach in conjunction with Gaussian Process regression, we achieved significant results with the estimated effects functions. As depicted in Figure~\ref{fig:case_study}, we successfully predicted the overall potential outcomes function and the spillover potential outcomes function, using experimental data from 33523 individuals among the entire population of 125286 individuals.

As shown in Figure~\ref{fig:case_study}, the results indicated that both the overall treatment effects function and the spillover effects function increase at the beginning and then decrease after reaching the maximum. This aligns with psychological patterns observed in gaming, where having a few treated friends might encourage one's increased engagement in the game. However, if too many friends receive treatment, it can lead to a decrease in an individual's effect due to a lack of perceived exclusivity or feelings of jealousy. Moreover, we observe the peak of the overall treatment effects occurs to the right of the spillover effects, suggesting that individuals who are treated themselves are more tolerant of a larger proportion of their friends being treated. Additionally, the small difference between the overall treatment effect and spillover effect at the endpoints when \(G=0\) or \(G=1\) reflects that when all or none of an individual's friends are treated, the individual's own treatment has minimal effects.  This is consistent with the inherent social nature of the game. 

Based on the results, we identified the treatment level that yields the maximum effect \((A = 1, G = 0.612)\). However, it is not feasible to allocate everyone in the population to this treatment level \((A = 1, G = 0.612)\). While determining the optimal decision-making strategy based on the estimated functions is not the primary focus of this paper, the results roughly suggest that treating about 50\% of the individuals in the population could potentially achieve the optimal average treatment effect.

Additionally, we conducted an analysis under the assumption of ignoring interference. This resulted in two horizontal lines in \cref{fig:case_study}. We found that the average treatment effect, while ignoring interference, is a negative value, \(-0.673\). In contrast, when considering interference, the optimal average treatment effect at \((A = 1, G = 0.612)\) is a positive value, \(2.681\). This indicates that ignoring interference in the analysis could mislead decision-making, suggesting a complete rejection of the treatment, which is not the case in reality. This discrepancy arises because the distribution of \(G\) in the population used for the experiment tends to favor smaller values, where the direct treatment effects are negative, leading to a negative average treatment effect when interference is ignored. This underscores the necessity and advantage of considering interference, as emphasized by our proposed method.

\section{Conclusion}
\label{sec:conclusion}

We propose a novel active learning approach that enables data-driven experimental design, facilitating causal research and estimation of treatment effects in extensive network data with interference, using minimal data. Our method, validated through simulation experiments, demonstrates the feasibility of obtaining accurate effects estimation with reduced data requirements, significantly enhancing data efficiency. Additionally, a real-world data application has confirmed our approach's practical usability and value in actual business contexts.

Looking forward, potential innovations could involve increasing network complexity beyond the static network structures used in this paper. In real-world settings, dynamic network structures are common, where relationships might exist only temporarily, and individuals might have different neighborhoods at varying times. Moreover, the current methodology for integrating neighbors' covariates and treatments may not be the most effective way to extract information from neighborhoods. Advanced techniques like neural networks or other dimensionality reduction methods might be more efficient to extract information from neighborhoods. Moreover, regarding the current method of integrating neighbors' treatments, within a network, the states \((a=0, g=0)\) and \((a=1, g=1)\) are easily achievable. However, reaching states like \((a=0, g=1)\) and \((a=1, g=0)\) can be challenging. If a sufficient number of these two types of data is required, extensive experimentation across multiple networks may be necessary. Perhaps a method could be developed, for instance, by using the easily attainable states \((a=0, g=0)\) and \((a=1, g=1)\), or by combining other readily available treatment level scenarios. This could enable the estimation of outcomes under challenging treatment levels like \((a=0, g=1)\) and \((a=1, g=0)\) without the need for direct experimentation in these hard-to-achieve conditions. Furthermore, guided by our estimated functions, decision-making can be informed to potentially determine what percentage of individuals in the population should be treated, or to devise an optimal vector-based treatment assignment for a group within a network.

\bibliographystyle{ACM-Reference-Format}
\bibliography{sample-base}

\appendix

\section{Appendix}

\subsection{Combined kernel function}
\label{subsec:Combined-kernel-function}

In order to perform Gaussian Process Regression (GPR), we need to define our kernel to represent the correlation between two data points. In our GPR problem, the features are of two types: one is the integrated neighbors' treatment \( G \), and the other is the integrated covariates \( X \). In the causal structure, these two types of variables have distinct properties; hence, they should not be considered equivalent in the kernel. For instance, the kernel function is employed to model the influence of neighbors' treatments and the interaction between this treatment and covariates on the potential outcome (\cite{rasmussen2006gaussian}),
\begin{align}
k((\boldsymbol{x}, g), (\boldsymbol{x'}, g')) = k_g(g,g') + \sum_{j=1}^{p} k_j((x_j,g),(x_j',g'))
\end{align}
Alternatively, with a primary focus on scenarios where the covariates \( X \) are high-dimensional, and recognizing that different types of covariates exert anisotropic effects on the outcome, we have developed a specific kernel to tackle this issue. This kernel is intentionally designed to mirror the distinct causal relationships between \( G \) and \( X \), and to effectively capture the anisotropic nature inherent within \( X \),
\begin{align}
k((x, g), (x', g')) = \, &\sigma_x^2 \exp \left( -\frac{1}{2} (x - x')^T Q (x - x') \right) \\
&+ \sigma_g^2 \exp \left( -\frac{1}{2 \lambda_g^2} (g - g')^2 \right) \nonumber
\end{align}
Where \(Q\) is a positive definite, diagonal matrix, borrowing idea from \cite{rasmussen2006gaussian} and \cite{vivarelli1998discovering}.

\subsection{Optimizing hyper-parameters}
\label{subsec:Optimizing-hyper-parameters}

To optimize the hyperparameters such as the length-scale and signal variance in Gaussian Process Regression (GPR), one typically maximizes the marginal (log) likelihood function of the training data (\cite{schulz2018tutorial,rasmussen2006gaussian}). 

Given a set of training data and hyper-parameters denoted as \(\boldsymbol{\theta}\), the log marginal likelihood is expressed as:
\[
\begin{aligned}
\label{eq:log-likelihood}
\log p(\boldsymbol{y_{t}} | \boldsymbol{X}, \boldsymbol{g}, \boldsymbol{\theta}) = &-\frac{1}{2} \boldsymbol{y_{t}^T} [\boldsymbol{K_t} + \sigma_t^2\boldsymbol{I}]^{-1} \boldsymbol{y_{t}} - \frac{1}{2} \log | \boldsymbol{K_t} + \sigma_t^2\boldsymbol{I}| \\
& - \frac{n_t}{2} \log 2\pi
\end{aligned}
\]

To utilize gradient descent methods such as Standard Gradient Descent, Conjugate Gradient Descent, and Limited-memory Broyden–Fletcher–Goldfarb–Shanno Algorithm, it's necessary to calculate the partial derivatives of the log likelihood function Equation~\ref{eq:log-likelihood} with respect to the hyper-parameters \(\boldsymbol{\theta}\). The derivative can be expressed as:
\[
\begin{aligned}
\frac{\partial \log p(\boldsymbol{y_{t}} | \boldsymbol{X}, \boldsymbol{g}, \boldsymbol{\theta})}{\partial \theta_j} 
&= \frac{1}{2}\boldsymbol{y_{t}^T} [\boldsymbol{K_t} + \sigma_t^2\boldsymbol{I}]^{-1} \boldsymbol{y_{t}} \\
&\quad- \frac{1}{2} \text{tr}\left([\boldsymbol{K_t} + \sigma_t^2\boldsymbol{I}]^{-1} \frac{\partial [\boldsymbol{K_t} + \sigma_t^2\boldsymbol{I}]}{\partial \theta_j}\right) \\
&= \frac{1}{2} \text{tr}\left(\left(\boldsymbol{\alpha} \boldsymbol{\alpha^T} - [\boldsymbol{K_t} + \sigma_t^2\boldsymbol{I}]^{-1} \right) \frac{\partial [\boldsymbol{K_t} + \sigma_t^2\boldsymbol{I}]}{\partial \theta_j}\right)
\end{aligned}
\]
with \(\alpha = [\boldsymbol{K_t} + \sigma_t^2\boldsymbol{I}]^{-1} \boldsymbol{y_{t}}\).

\subsection{Genetic Algorithm}
\label{subsec:genetic-algorithm}

The \cref{alg:optimization_network} presents a simplified version of the genetic algorithm used in our computations. In practice, to prevent overfitting, an early stopping criterion is implemented, which terminates the algorithm if there is no improvement in the fitness value within a certain number of generations. Additionally, the initialization function, roulette-wheel selection function, and crossover mutation function utilized within the algorithm have been specifically tailored to address our particular problem set.
\begin{algorithm}
   \caption{Optimization for Current Network}
   \label{alg:optimization_network}
\begin{algorithmic}
   \STATE {\bfseries Input:} target neighbors' treatment \(g\), target's own treatment \(a\), blocking length \(\alpha\), relation matrix \(W\), number of epochs \(\text{epoch}\), parameter \(k\), number of batches per epoch \(K\)
   \STATE Initialize the initial set of \(K\) parents \(\boldsymbol{\mathcal{A}^K}\)
   \FOR{each epoch \(i\)}
       \STATE Calculate fitness value \(F^K\) for the \(K\) parent vector-based treatments \(\boldsymbol{\mathcal{A}^K}\) using \cref{eq:fit}
       \FOR{\(i\) in \(K\)}
           \STATE Select two parents from the \(K\) parents using a roulette-wheel selection method
           \STATE Generate two children by crossover of the two selected parents, and calculate the fitness value for each child using \cref{eq:fit}; select the child with the higher fitness value
           \STATE Store this selected child in \(\boldsymbol{\mathcal{A}_C^K}\)
       \ENDFOR
       \STATE From \(\boldsymbol{\mathcal{A}^K}\) and \(\boldsymbol{\mathcal{A}_C^K}\), select \(K\) treatments with the higher fitness value to update \(\boldsymbol{\mathcal{A}^K}\)
   \ENDFOR
   \STATE \textbf{return} The best treatment assignment is the one in \(\boldsymbol{\mathcal{A}^K}\) with the highest fitness value
\end{algorithmic}
\end{algorithm}

\subsection{Random treatment allocation}
\label{subsec:Random-treatment-allocation}

Similar to the proposed method, the random treatment allocation also initiates with base points \(A = 1, G = 1\) and \(A = 0, G = 0\). Then, the selection process alternates between points with \(A = 1\) and \(A = 0\), striving for a balance in representation. However, we decide \( g^* \) randomly from a uniform distribution over the interval (0, 1), denoted as \( g^* \sim \mathcal{U}(0, 1) \). After determining \( g^* \), to ensure comparability with the guidelines of the proposed method, it is necessary to ensure that the simulation of the random treatment allocation method uses the same number of networks as the proposed method. Assume that the proposed method uses \(N_u\) networks and predicts effect values for \(n_u\) \((A, G)\) point pairs. The number of networks used for treatment assignments in the random treatment allocation can be represented as \( \boldsymbol{N_r} \), a vector of length \(n_u\), as seen in Algorithm~\ref{alg:even_distribution}. Subsequently, for nodes of selected networks, we assign \(A = 1\) to a proportion of nodes as determined by \(g^*\).

\begin{algorithm}[H]
\caption{Even Distribution of an Integer}
\label{alg:even_distribution}
\begin{algorithmic}
\REQUIRE Integer $N_u$, number of parts $n_u$
\ENSURE Vector $\boldsymbol{N_r}$ of length $n_u$ with distributed values
\STATE $base \gets \lfloor N_u / n_u \rfloor$
\STATE $remainder \gets N_u \mod n_u$
\STATE Initialize vector $\boldsymbol{N_r}[1\ldots n_u]$ to $base$
\FOR{$i = 1$ to $remainder$}
    \STATE $\boldsymbol{N_r}[i] \gets \boldsymbol{N_r}[i] + 1$
\ENDFOR
\RETURN $\boldsymbol{N_r}$
\end{algorithmic}
\end{algorithm}

\end{document}